\documentclass[osajnl,twocolumn,showpacs,superscriptaddress,10pt]{revtex4-1} 
\usepackage{amsmath,amssymb,graphicx,color}

\definecolor{red}{rgb}{0.9,0,0}
\definecolor{green}{rgb}{0,0.8,0}
\definecolor{blue}{rgb}{0,0,0.8}
\definecolor{cautionred}{rgb}{1.0,0,0}

\definecolor{maroon}{rgb}{0.7,0,0}

\definecolor{ngreen}{rgb}{0.3,0.7,0.3}

\definecolor{golden}{rgb}{0.8,0.6,0.1}

\begin{document}

\title{Analog of the Clauser-Horne-Shimony-Holt inequality for steering}

\author{Eric G. Cavalcanti} 

\affiliation{School of Physics, The University of Sydney, Sydney, NSW 2006, Australia}

\author{Christopher J. Foster}

\affiliation{School of Mathematics and Physics, The University of Queensland, 
Brisbane, QLD 4072, Australia}

\author{Maria Fuwa}

\affiliation{Department of Applied Physics, School of Engineering, The University of Tokyo,
7-3-1 Hongo, Bunkyo-ku, Tokyo 113-8656, Japan}

\author{Howard M. Wiseman}

\affiliation{Centre for Quantum Computation and Communication Technology 
(Australian Research Council), Centre for Quantum Dynamics, Griffith University, Brisbane, QLD 4111, Australia}

\begin{abstract}
The Clauser-Horne-Shimony-Holt (CHSH) inequality (and its permutations), are necessary and sufficient criteria for Bell nonlocality in the simplest Bell-nonlocality scenario: 2 parties, 2 measurements per party and 2 outcomes per measurement. Here we derive an inequality for EPR-steering that is an analogue of the CHSH, in that it is necessary and sufficient in this same scenario. However, since in the case of steering the device at Bob's site must be specified (as opposed to the Bell case in which it is a black box), the scenario we consider is that where Alice performs two (black-box) dichotomic measurements, and Bob performs two mutually unbiased qubit measurements. We show that this inequality is strictly weaker than the CHSH, as expected, and use it to decide whether a recent experiment [Phys. Rev. Lett. 110, 130401 (2013).] involving a single-photon split between two parties has demonstrated EPR-steering.
\end{abstract}


\maketitle 

\section{Introduction}

In Ref.~\citep{Wiseman2007}, one of us and coworkers introduced a new classification of quantum nonlocality, a formalisation of the concept of \emph{steering} introduced by Schr\"odinger in 1935  \cite{Schrodinger1935} as a generalization of the Einstein-Podolsky-Rosen (EPR) paradox~\cite{EPR1935}. Steering can be understood as the failure of a hybrid \emph{Local 
Hidden Variable} (LHV)--\emph{Local Hidden State} (LHS) model to reproduce the correlations between two subsystems. Here, `state' means a quantum state for one subsystem (traditionally Bob's), 
as opposed to the `variables' used to describe Alice's system which are not constrained to correspond to a quantum state. Apart from assuming local causality in the same sense as Bell used, an LHV--LHS model assumes that one of the parties (Bob) performs trusted quantum measurements on his subsystem, whereas the other (Alice) does not trust her measurement apparatus. Equivalently, it can be thought of as a model where Bob is in possession of a system in some unknown quantum state, only classically correlated with some arbitrary variables producing the outcomes at Alice's site. 

In Ref.~\cite{Cavalcanti2009}, two of us and coworkers introduced the concept of \textit{EPR-steering inequalities} as any criterion which demonstrates that a set of correlations observed by Alice and Bob cannot be described by an LHV--LHS model. The first example of such inequalities, based upon multiplying conditional variances, was introduced  in the seminal paper  by Reid in 1989~\cite{Reid1989,ReidReview2009}, to create an experimentally accessible version of the EPR-paradox, applicable to continuous variables systems, as in EPR's original argument. An inequality for the case of discrete variables, as discussed in Bohm's version of the EPR paradox, was introduced by one of us and coworkers in \cite{Cavalcanti2007}. That these {\it Reid inequalities}, as they are sometimes known, can be derived from the LHV--LHS model, was shown in Ref.~\cite{Cavalcanti2009}. EPR-steering has been demonstrated over macroscopic distances using both continuous variables~\cite{OuPRL1992, BowenPRL2003, HowelPRL2004, ReidReview2009, HandchenNatPh2012, Schneeloch2013} and discrete variables~\cite{SaundersNatPhys2010,SmithNatCom2012,BennetPRX2012,Wit12}. There have also been many recent theoretical advances, even for the simple case relevant to two qubits. This includes showing that violation of an EPR-steering inequality is a necessary condition for one-sided device-independent quantum key distribution~\cite{Bra12}, determining the most parsimonious demonstration~\cite{Parsim2012}, showing that one-way EPR-steerability exists for two-qubit states~\cite{Bowles}, and relating EPR-steerability to the so-called steering ellipsoid~\cite{Sania2015}.

In this paper we present a new theoretical result: an EPR-steering inequality that is necessary and sufficient for a set of correlations involving two settings and two outcomes per site, with mutually unbiased measurements at Bob's site. By applying this inequality, we then put to test, for the first time, whether CHSH-type~\cite{Clauser1969} correlations observed in a recent experiment~\cite{MorinPRL13} exhibit EPR-steering. The interesting thing about this experiment was that the entanglement was provided by a single-photon split between two parties. We stress that the authors of Ref.~\cite{MorinPRL13}, although using  CHSH-type correlations, are clear that they do not demonstrate Bell-nonlocality, because of the nonidealities of their measurements; what they do demonstrate is entanglement, with minimal assumptions about the state. Nevertheless their detection efficiencies were quite high, so it certainly seems possible that their experiment could have been a demonstration of EPR-steering. They do not consider this question in their paper, and it is only with the theory presented here that it is clear how to answer the question definitively. We conclude with open questions.

\section{Locality models}

Let us introduce our notation. Given a pair of systems at Alice and Bob, denote $\mathfrak{D}_{\alpha}$ and $\mathfrak{D}_{\beta}$ the sets of observables in the Hilbert space of Alice's and Bob's system, respectively. An element of $\mathfrak{D}_{\alpha}$ is denoted by $A$, with a set of outcomes labeled by $a\in \mathfrak{L}(A)$, and similarly for Bob. The joint state $W$ of the system is steerable by Alice iff it is \emph{not} the case that for all $a\in\mathfrak{L}(A)$, $b\in \mathfrak{L}(B)$, $A\in\mathfrak{D}_{\alpha}$, $B\in\mathfrak{D}_{\beta}$, the joint probability distributions can be written in the form
\begin{equation}
P(a,b|A,B;W)=\sum_{\lambda}\wp(\lambda)\wp(a|A,\lambda)P(b|B;\rho_{\lambda}),\label{eq:no-steering}
\end{equation}
where $\wp(a|A,\lambda)$ denotes an arbitrary probability distribution and $P(b|B;\rho_{\lambda})$ denotes the quantum probability of outcome $b$ given measurement $B$ on state \textbf{$\rho_{\lambda}$}.

This can be understood as follows: if the state is not steerable, then the correlations between Alice and Bob could be generated by a distribution of local variables $\lambda$, each of which determines arbitrarily the probabilities for outcomes at Alice's side and a quantum state on Bob's side. For this reason Wiseman \emph{et al.} called the model of (\ref{eq:no-steering}) a local hidden state (LHS) model \citep{Wiseman2007}, though here we adopt the more transparent term, an {LHV--LHS} model, as explained in the introduction. The question we will address is: if Alice and Bob each have a choice of two dichotomic observables, how can we prove that their experimental outcomes cannot be explained by that model?

We thus consider a scenario where Alice and Bob perform subsets $\mathcal{\mathfrak{M}}_{\alpha}\subseteq\mathfrak{D}_{\alpha}$ and $\mathcal{\mathfrak{M}}_{\beta}\subseteq\mathfrak{D}_{\beta}$ of all observables, called \emph{measurement strategies}. The correlations $P(a,b|A,B)$ for a measurement strategy $\mathcal{\mathfrak{M}}_{\alpha}$ and $\mathfrak{M}_{\beta}$ exhibit steering if and only if it is not the case that there exists a model of type (\ref{eq:no-steering}) for all $a\in\mathfrak{L}(A)$, $b\in\mathfrak{L}(B)$, $A\in\mathcal{\mathfrak{M}}_{\alpha}$ and $B\in\mathfrak{M}_{\beta}$.  Obviously, if for a given measurement strategy the correlations have an {LHV--LHS} model, this does not imply that the underlying \emph{state $W$} is not steerable, since there could be another strategy that does not. However, we can still ask the question of whether there are necessary and sufficient conditions for a particular measurement strategy to demonstrate steering.

In \citep{Wiseman2007}, it was shown that steering is a form of nonlocality strictly weaker than Bell nonlocality. That is, a state has no {LHV--LHS} model if (but not only if) it has no LHV model (i.e.~an LHV model for both parties). The correlations between Alice and Bob are said to have an LHV model when for all $a\in\mathfrak{L}(A)$, $b\in\mathfrak{L}(B)$, $A\in\mathfrak{D}_{\alpha}$, $B\in\mathfrak{D}_{\beta}$, the joint probability distributions can be written in the form
\begin{equation}
P(a,b|A,B)=\sum_{\lambda}\wp(\lambda)\wp(a|A,\lambda)\wp(b|B,\lambda). \label{eq:LHV}
\end{equation} 
In that case, given a finite number of settings per site and a finite number of outcomes per setting, it is well-known that the set of correlations allowed by LHV theories live in a convex polytope -- a convex set the extreme points of which are the classical ``pure'' states, i.e., the states with well-defined values for all observables. The Bell inequalities are the linear inequalities that define the facets of this polytope, and therefore they completely characterise the set. Given the set of experimental probabilities $\wp(a,b|A,B)$ for all $a,b,A,B$, this has an LHV model if and only if it satisfies all Bell inequalities for that number of settings and outcomes. In the simplest Bell scenario, involving two observers with 2 dichotomic measurements per site, the CHSH inequality \cite{Clauser1969} and its permutations define all the non-trivial facets of the LHV polytope. Here we ask whether a similar inequality can be derived for the same scenario. This is the the simplest steering scenario involving projective measurements, although a simpler demonstration is possible if Bob uses a single trichotomic positive-operator-valued measure~\cite{Parsim2012}, in the sense defined in that reference.

The set of correlations that have an {LHV--LHS} model for a given scenario also form a convex set~\cite{Cavalcanti2009}, and we will now proceed to study that set. In particular, we can rewrite any {LHV--LHS} model in terms of a model that involves a distribution only over the extreme points of that convex set. That is, we can simplify equation (\ref{eq:no-steering}) by noting that we can always decompose $\wp(a|A,\lambda)P(b|B;\rho_{\lambda})$ into $\sum_{\chi}\int d\xi\wp(\chi,\xi|\lambda)\delta_{a,f(A,\chi)}P(b|B;|\psi_{\xi}\rangle\langle\psi_{\xi}|)$, where $\chi$  are variables which determine all values of the observables $A$ via the function $f(A,\chi)$, and $\xi$ determines a pure state $|\psi_{\xi}\rangle$ for Bob. Thus if $\Pi_{b}^{B}$ is the projection operator corresponding to outcome $b$ of measurement $B$, equation (\ref{eq:no-steering}) becomes (where we omit reference to $W$ since we make no assumption about the state that produces the correlations) 
\begin{equation}
P(a,b|A,B)=\sum_{\chi,\xi} \wp(\chi,\xi) \delta_{a,f(A,\chi)} \langle \psi_\xi | \Pi_{b}^{B} | \psi_\xi \rangle.
\label{eq:no-steering extreme}
\end{equation}

\section{EPR-steering with two dichotomic measurements per site}

We will now focus on the case where Alice and Bob each have a choice between two dichotomic measurements to perform: \{$A$, $A'$\}, \{$B$,$B'$\}. The outcomes of $A$ will be labelled $a\in\{-1,1\}$ and similarly for the other measurements.

The extremal values for Alice and Bob can each be written in a vector form. For Alice we define the vector 
\begin{equation}
\vec{P}_{\mathcal{A}}(\chi)=(p_{1}^{A}(\chi),p_{-1}^{A}(\chi),p_{1}^{A'}(\chi),p_{-1}^{A'}(\chi))^{\rm T},\label{eq:Pa}
\end{equation}
where $p_{1}^{A}(\chi) = \delta_{1,f(A,\chi)}$, etc. Obviously there are only four distinct such extremal vectors. For Bob we define a probability vector $\vec{P}_{\mathcal{B}}(\xi)$ similarly, but as a function of the hidden variable $\xi$, and where $p_{b}^{B}(\xi)= \langle \psi_\xi | \Pi_{b}^{B} | \psi_\xi \rangle .$ Of course there is a continuous infinity of such vectors for Bob. We call $\vec{P}_{\mathcal{A}}(\chi) \vec{P}_{\mathcal{B}}^{\rm T}(\xi)$ an extremal matrix. It generates the {\it correlation matrix} $\mathbf{P}_{\mathcal{AB}}$, which encodes the joint probabilities for all combinations of measurements via 
\begin{equation}
\mathbf{P}_{\mathcal{AB}}=\sum_{\chi}\int d\xi\wp(\chi,\xi)\vec{P}_{\mathcal{A}}(\chi) \vec{P}_{\mathcal{B}}^{\rm T}(\xi).\label{eq:P}
\end{equation}
where $\sum_{\chi}\int d\xi\wp(\chi,\xi)=1$. The $16$ elements of this matrix are all the distinct $P(a,b|A,B)$, which comprise all the data that Alice and Bob extract from their measurements. Our problem
can now be rephrased as whether one can, given the experimental $\mathbf{P}_{\mathcal{AB}}$, decide whether it can be written as (\ref{eq:P}).

As an example, here is the  extremal matrix for $\vec{P}_{\mathcal{A}}(\chi)=(1,0,1,0)^{\rm T}$ and an arbitrary state $|\psi_{\xi}\rangle$:
\[
\left[\begin{array}{cccc}
p_{1}^{B}(\xi) & 0 & p_{1}^{B}(\xi) & 0\\
1-p_{1}^{B}(\xi) & 0 & 1-p_{1}^{B}(\xi) & 0\\
p_{1}^{B'}(\xi) & 0 & p_{1}^{B'}(\xi) & 0\\
1-p_{1}^{B'}(\xi) & 0 & 1-p_{1}^{B'}(\xi) & 0
\end{array}\right].
\]
The matrices for the other three extremal vectors at Alice's side are similar to this, but with some columns permuted. Since Bob's probabilities arise from measurements on a quantum system, there are constraints that they must obey, such as uncertainty relations. We will now study those constraints.

Let the basis of eigenstates for Bob's observable $B$ be labelled as $\{|1\rangle,|-1\rangle\}$. The projector for outcome $1$ of $B'$ can  be parametrized by $\mu$ and $\phi$ as follows  
\begin{eqnarray}
\Pi_{1}^{B'} & = & (\sqrt{\mu}\,|1\rangle+\sqrt{1-\mu}\, e^{i\phi}|-1\rangle) \nonumber\\
 &  & \qquad\times(\sqrt{\mu}\,\langle1|+\sqrt{1-\mu}\, e^{-i\phi}\langle-1|)\,.
\end{eqnarray}
Similarly, we can write an arbitrary pure state as 
\begin{equation}
|\psi_{\mu',\phi'}\rangle=\sqrt{\mu'}\,|1\rangle+\sqrt{1-\mu'}\, e^{i\phi'}|-1\rangle.
\end{equation} 
Then 
\begin{eqnarray}
p_{1}^{B}(\mu',\phi')&\equiv &\langle\psi_{(\mu',\phi')}|\Pi_{1}^{B}|\psi_{(\mu',\phi')}\rangle=\mu', \nonumber \\
p_{1}^{B'}(\mu',\phi')&=&\mu\mu'+ (1-\mu)(1-\mu') \nonumber \\ 
&& + 2\sqrt{\mu(1-\mu)\mu'(1-\mu')}\, \cos(\phi'-\phi).
\end{eqnarray} 

\begin{figure}
\includegraphics[scale=0.6]{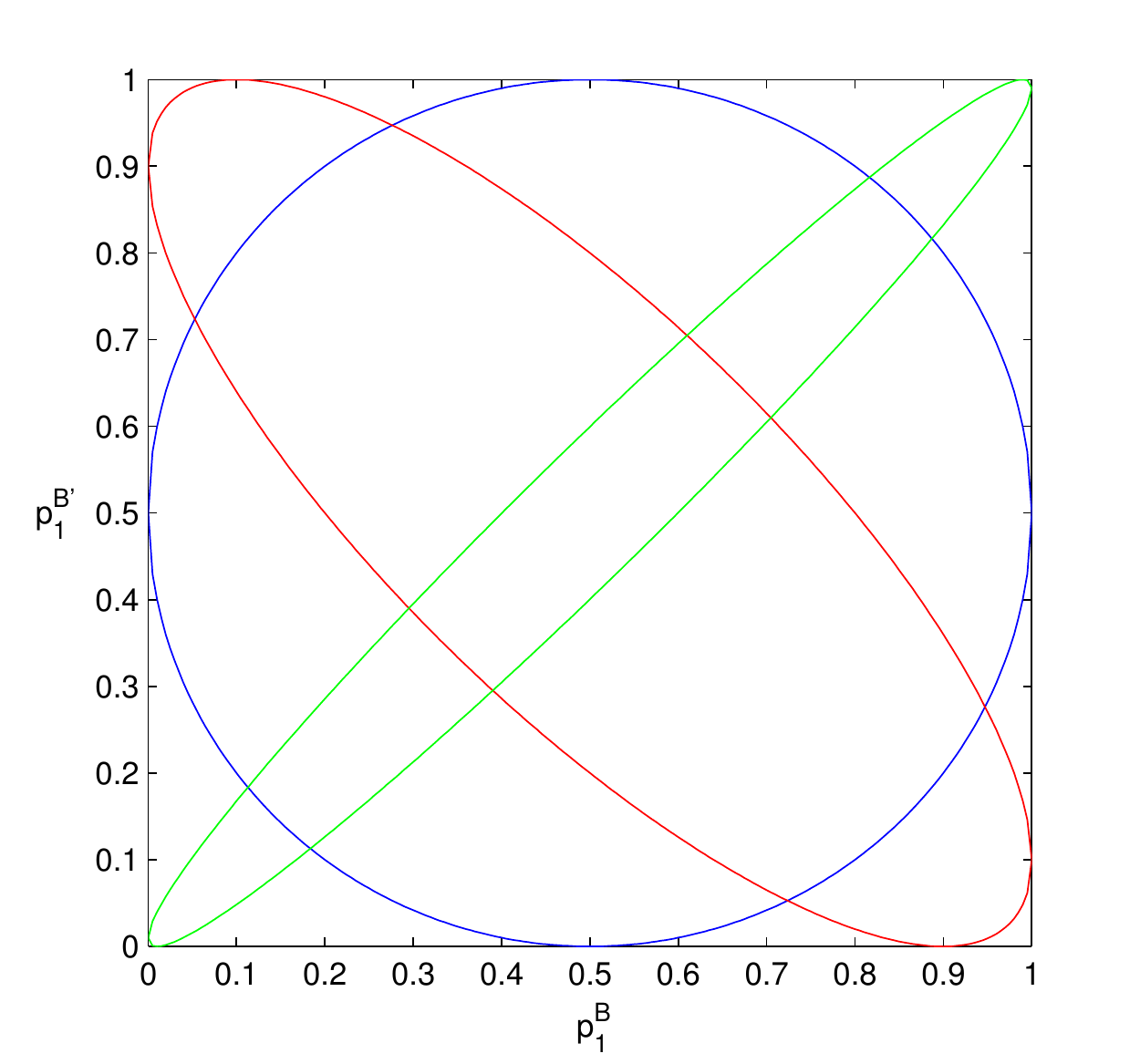}
\protect\caption{The boundaries of the sets of allowed quantum probabilities $(p_{1}^{B},p_{1}^{B'})$ for outcomes associated with two projective measurements $\Pi_{1}^{B}$, $\Pi_{1}^{B'}$ on a qubit are ellipses defined by the parameter $\mu=\mathrm{Tr}\{\Pi_{1}^{B}\Pi_{1}^{B'}\}$. Here they are shown for $\mu=0.1$ (red), $\mu=0.5$ (blue), $\mu=0.99$ (green). \label{fig:convex_sets}} 
\end{figure}

Analysing the equations above, we find that the set of allowed quantum probabilities, i.e. the possible pairs of $(p_{1}^{B},p_{1}^{B'})$, form the convex hull of an ellipse with eccentricity characterised by the parameter $\mu=\mathrm{Tr}\{\Pi_{1}^{B}\Pi_{1}^{B'}\}$, which depends only on the measurements being performed, as shown in Fig.~\ref{fig:convex_sets}. The boundaries are achieved with $\mathrm{cos}(\phi'-\phi)=\pm1$. This completely characterises the set of allowed $\vec{P}_{\mathcal{B}}(\xi)$. Since we are interested in the extreme points of this set, we can reparametrize the ellipse using the real parameter $\xi$ to stand for any quantum state that can generate that point for a given value of $\mu$: 
\begin{eqnarray}\label{eq:ellipse}
p_{1}^{B}(\xi)-\frac{1}{2} & = & \frac{1}{2}[\sqrt{\mu}\,\cos(\xi)-\sqrt{1-\mu}\,\sin(\xi)]  \nonumber \\
p_{1}^{B'}(\xi)-\frac{1}{2} & = & \frac{1}{2}[\sqrt{\mu}\,\cos(\xi)+\sqrt{1-\mu}\,\sin(\xi)],\label{eq:parametrisation}
\end{eqnarray}
which reduces to a circle for $\mu=1/2$ and to diagonal or off-diagonal lines for $\mu=\pm1$. If one does not include the information about which observables are being measured, then the extreme points become the vertices of the square and we obtain for $\mathbf{P}_{\mathcal{AB}}$ the whole set of Bell-local correlations, the (non-trivial) boundaries of which are given by the CHSH inequalities.

Looking at an arbitrary $\mathbf{P}_{\mathcal{AB}}$ representing statistical data from Alice's and Bob's measurements, we find that it has to satisfy certain constraints. Four of these are normalisation conditions of the form $\sum_{a,b}P(a,b|A,B)=1$. Another 8 constraints come from non-signalling conditions of the form $\sum_{a}P(a,b|A,B)=\sum_{a'}P(a',b|A',B)=P(b|B)$. Therefore the number of free parameters in $\mathbf{P}_{\mathcal{AB}}$ is reduced from 16 to 4. It is sufficient then to consider the space of 4 parameters given by the correlations of the form 
\begin{equation}
\left\langle AB\right\rangle  =  P(a=b|A,B)-P(a=-b|A,B)\,.
\end{equation}
These correlations will have an LHV--LHS model if and only if they can be written in the form
\begin{equation}
\left\langle AB\right\rangle = \sum_{\chi}\int d\xi\wp(\chi,\xi)(2p_{1}^{A}(\chi)-1)(2p_{1}^{B}(\xi)-1)\,.
\end{equation}
Given the 4 correlations of form $\left\langle AB\right\rangle $ (and the constraints above), $\mathbf{P}_{\mathcal{AB}}$ is uniquely determined, and vice-versa. Therefore,  if there is an LHV--LHS model of the form (\ref{eq:P}) for $\mathbf{P}_{\mathcal{AB}}$ then there is an LHV--LHS model for this set of correlations, and vice-versa.

Let us look at the case $\mu=1/2$, which corresponds to spin measurements along orthogonal spatial directions on Bob's system. Then the boundary is a circle and (\ref{eq:parametrisation}) can be reparametrized as (we will keep the label $\xi$ for notational simplicity but note that it is different from that of Eq.~\ref{eq:ellipse}) 
\begin{eqnarray}
2p_{1}^{B}(\xi)-1 & = &\cos(\xi)\nonumber \\
2p_{1}^{B'}(\xi)-1 & = &\sin(\xi).\label{eq:parametrisation for p=00003D1/2}
\end{eqnarray}
Now we will label $\chi\in\{1,2,3,4\}$, corresponding respectively to $(p_{1}^{A}=p_{1}^{A'}=1)$, $(p_{1}^{A}=1,p_{1}^{A'}=0)$, $(p_{1}^{A}=0,p_{1}^{A'}=1)$ and $(p_{1}^{A}=p_{1}^{A'}=0)$. We can thus see that for each value of $(\chi,\xi)$ the four correlation functions can be written as
\begin{equation}
\begin{array}{ccccc}
 & \chi=1 & \chi=2 & \chi=3 & \chi=4 \\
\left\langle AB\right\rangle (\xi) &\cos(\xi) &\cos(\xi) & -\cos(\xi) & -\cos(\xi)\\
\left\langle A'B\right\rangle (\xi) &\cos(\xi) & -\cos(\xi) &\cos(\xi) & -\cos(\xi)\\
\left\langle AB'\right\rangle (\xi) &\sin(\xi) &\sin(\xi) & -\sin(\xi) & -\sin(\xi)\\
\left\langle A'B'\right\rangle (\xi) &\sin(\xi) & -\sin(\xi) &\sin(\xi) & -\sin(\xi)
\end{array}\label{eq:extreme points}
\end{equation}

The two last columns can be obtained from the first two by making $\xi\rightarrow\xi+\pi$, so the only relevant extreme points are $\chi=1$ and $\chi=2$. All points in the convex set of LHV--LHS correlations can be written as a convex combination of the first two columns for each value of $\xi$. What are the boundaries of that set?

We start with some partial answers to that question. Four inequalities can be immediately derived that must be satisfied by all points in the set of LHV-LHS correlations:
\begin{eqnarray}\label{eq:first_ineq}
&\left\langle AB\right\rangle ^{2}+\left\langle A'B'\right\rangle ^{2} & \leq 1 \nonumber \\
&\left\langle A'B\right\rangle ^{2}+\left\langle AB'\right\rangle ^{2} & \leq 1 \nonumber\\
&\left\langle AB\right\rangle ^{2}+\left\langle AB'\right\rangle ^{2} & \leq 1 \nonumber \\
&\left\langle A'B\right\rangle ^{2}+\left\langle A'B'\right\rangle ^{2} & \leq 1
\end{eqnarray}

We can also see that the LHV--LHS correlations must satisfy an inequality with the same form of the CHSH inequality. Clearly for each column of (\ref{eq:extreme points})
\begin{equation}
\left\langle AB\right\rangle +\left\langle A'B\right\rangle +\left\langle AB'\right\rangle -\left\langle A'B'\right\rangle \leq2\label{eq:CHSH}
\end{equation}
is satisfied, and therefore it must be also satisfied for the total correlations. 

Inequality \eqref{eq:CHSH} is, of course, satisfied by all LHV correlations. However, all inequalities in \eqref{eq:first_ineq} can attain the maximum algebraic value of 2 with LHV models (where we do not trust Bob's apparatuses and thus not require that the measurements performed by Bob correspond to the previously defined operators). In fact, there are no non-trivial bounds on LHV correlations with only two correlations, because in an LHV model the values of the variables in one correlation can be set independently of those in the other. 

Looking for quantum violations, we find that the last two inequalities in Eq.~\eqref{eq:first_ineq} cannot be violated, since they have the same choice of setting for Alice, and thus essentially reduce to a quantum uncertainty relation for Bob's measurements. The first two can attain the maximal algebraic value of 2 with a maximally entangled state and $A$ and $A'$ along the same direction as the corresponding operator for Bob within each correlation function. The CHSH inequality \eqref{eq:CHSH}, as is well-known, can attain a maximum value of $2\sqrt{2}$ with quantum correlations.

\section{Necessary and sufficient EPR-steering inequality}

As explained before, the set of LHV--LHS correlations for the scenario we are considering is defined by all convex combinations of vectors which can be written in the form of the first or second columns of \eqref{eq:extreme points}. We define the sets $\mathcal{C}_{1}$ and $\mathcal{C}_{2}$ by the convex hull of two sets of points in $\mathbb{R}^{4}$ parametrised by $\xi$: 
\begin{align}
\mathcal{C}_{1} & :=\mathrm{convex}\left(\left\{ (\cos\xi,\cos\xi,\sin\xi,\sin\xi)\colon\xi\in[0,2\pi]\right\} \right)\nonumber \\
\mathcal{C}_{2} & :=\mathrm{convex}\left(\left\{ (\cos\xi,-\cos\xi,\sin\xi,-\sin\xi)\colon\xi\in[0,2\pi]\right\} \right).\label{eq:sets_C1_C2_definitions}
\end{align}
The convex set $\mathcal{C}$ of LHV--LHS correlations is then the convex hull of the union of $\mathcal{C}_{1}$ and $\mathcal{C}_{2}$:
\begin{equation}
\mathcal{C}:=\mathrm{convex}(\mathcal{C}_{1}\cup \mathcal{C}_{2}).\label{eq:set_C_definition}
\end{equation}

The first step is to notice that $\mathcal{C}_{1}$ and $\mathcal{C}_{2}$ are in fact filled-in circles in orthogonal subspaces of $\mathbb{R}^{4}$. Define
a new orthogonal basis by the vectors 
\begin{align}
\mathbf{e}_{1} & =(1,1,0,0)\nonumber \\
\mathbf{e}_{2} & =(0,0,1,1)\nonumber \\
\mathbf{e}_{3} & =(1,-1,0,0)\nonumber \\
\mathbf{e}_{4} & =(0,0,1,-1)
\end{align}
In terms of this basis, the vectors which make up the boundary of $\mathcal{C}_{1}$ are of the form $\cos(\xi)\mathbf{e}_{1}+\sin(\xi)\mathbf{e}_{2}$, while the vectors making up the boundary of $\mathcal{C}_{2}$ are $\cos(\xi)\mathbf{e}_{3}+\sin(\xi)\mathbf{e}_{4}$. From now on, we write all vectors in terms of the basis $\{\mathbf{e}_{i}\}$.

As we will show, the following inequality describes the volume inside $\mathcal{C}$. Writing $\mathbf{v}=(v_{1},v_{2},v_{3},v_{4})$ in the basis $\{\mathbf{e}_{i}\}$, we have $\mathbf{v}\in \mathcal{C}$ if
and only if 
\begin{equation}
\sqrt{v_{1}^{2}+v_{2}^{2}}+\sqrt{v_{3}^{2}+v_{4}^{2}}\leq1\,.\label{eqn_ineq}
\end{equation}
This implies, once converted back into the original basis:
\begin{multline}
\sqrt{\left\langle (A+A')B\right\rangle ^{2}+\left\langle (A+A')B'\right\rangle ^{2}}\\
+\sqrt{\left\langle (A-A')B\right\rangle ^{2}+\left\langle (A-A')B'\right\rangle ^{2}}\leq2\,.\label{eq:ineq_nec_suf}
\end{multline}

This is our main result. Note that since all of the terms such as $\left\langle (A+A')B'\right\rangle ^{2}$ are positive semi-definite,
the left hand side of \eqref{eq:ineq_nec_suf} is always larger than or equal to the left hand side of the CHSH inequality \eqref{eq:CHSH} (or any of its permutations). Therefore the CHSH inequality is violated only if \eqref{eq:ineq_nec_suf}
is, i.e., the correlations violate local causality only if they demonstrate steering. And contrary to the inequalities in \eqref{eq:first_ineq}, \eqref{eq:ineq_nec_suf} uses all four correlations $\left\langle AB\right\rangle ,\,\left\langle A'B\right\rangle ,\,\left\langle AB'\right\rangle ,\,\left\langle A'B'\right\rangle $. 

Violation of this inequality is necessary and sufficient for demonstrating EPR-steering within this scenario. Proving that all points in $\mathcal{C}$ satisfy \eqref{eqn_ineq} is relatively easy; we do that first, followed by a proof that all points that satisfy \eqref{eqn_ineq} are in $\mathcal{C}$. These proofs can be found In Appendices \ref{appendix_A} and  \ref{appendix_B} respectively.

If Alice and Bob share a maximally entangled state and Alice performs projective measurements, the maximum violation of this inequality happens when Alice's measurements are mutually unbiased, when the expression on the left side of \eqref{eqn_ineq} becomes $2\sqrt{2}$. Interestingly, this value is independent of the relative angle between Alice's measurement and Bob's measurements, unlike the CHSH inequality, which requires a specific relation between those measurements, and unlike most known EPR-steering inequalities, with the exception of those in Ref.~\cite{Parsim2012}. The proof of these statements can be found in Appendix \ref{appendix_C}.

\section{Application to an Experiment}

\begin{figure}[!b]
\begin{center}
\includegraphics[bb=163 127 479 341, width=\linewidth]{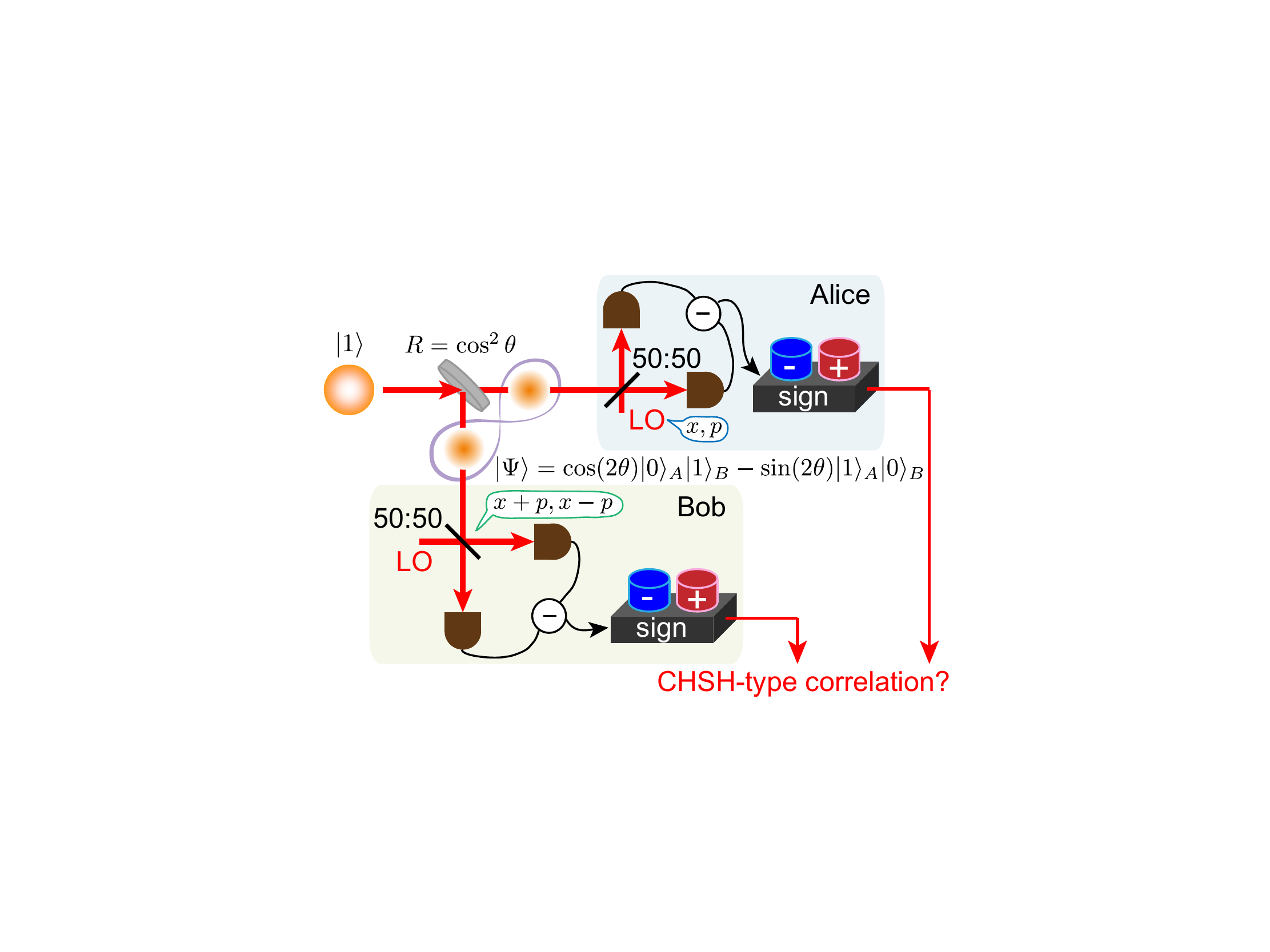}
\end{center}
\vspace{-4mm}
\caption{Experimental setup in Ref.~\cite{MorinPRL13}. A single photon is split between 2 parties (Alice and Bob) using a beam splitter of reflectivity $R = \cos^2 \theta$. Both perform sign-binned homodyne measurements for the $\{ \hat{x}, \hat{p} \}$ and $\{ \hat{x} + \hat{p}, \hat{x} - \hat{p} \}$ quadratures respectively. The CHSH-type correlation is tested for the measurement outcomes. In contrast to this scheme which uses only 2 difference bases, in Ref.~~\cite{Fuwa2015}, Alice uses 6 difference bases, and Bob performs quantum tomography for his local quantum state to verify EPR-steering. }
\label{fig:exp_schematic}
\end{figure}

In this section we apply the above conditions to an experiment involving a single photon split between two parties, designed to ``witness trustworthy single-photon entanglement''~\cite{MorinPRL13} (Fig.~\ref{fig:exp_schematic}). In this experiment, because there is (to a good approximation) at most one photon, initially in a single mode, the entangled state is a two-qubit state. Moreover, the two parties measure a CHSH-type correlation, each using two measurements with binary outcomes. Although Bob's measurements are not actually projective, all of the correlations of his results with Alice's results are, as we will show, reduced below the correlations that would pertain to projective measurements by a simple multiplicative constant. Thus, with an adjustment for this multiplicative constant, we can apply our theory above to the measured correlations to discover whether this experiment amounts to a demonstration of EPR-steering or not.

\subsection{Modelling sign-binned homodyne measurements on a single photon}

A single-photon state, split between Alice and Bob, can be represented in the Fock basis as: 
\begin{equation}
| \Psi \rangle  = \cos(2\theta) |0 \rangle_A |1 \rangle_B - \sin(2\theta) |1 \rangle_A |0 \rangle_B ,
\label{eq:model_arbitrary_qubit}
\end{equation}
where $\theta$ parameterizes the splitting between the two parties. Restricted to the single-photon subspace, this is a two-qubit state as considered in the above theory. This is clearly an idealization. However the dominant deviation from the pure two-qubit state above in Ref.~\cite{MorinPRL13} is due to loss of the photon, or failure to generate it in the first place. That is, the state can more accurately be described as 
\begin{equation}
\rho = p_0 |0,0 \rangle_{AB} \langle 0,0 |_{AB} + p_1 | \Psi \rangle \langle \Psi | 
\end{equation}
where $p_1 = 1-p_0$ is the probability that one photon is present. This mixed state is still a two-qubit state, so we can still apply our theory. 

Turning now to the measurements performed in Ref.~\cite{MorinPRL13}, they are sign-binned homodyne measurements. For a mode containing at most one photon, the positive-operator-valued measure for homodyne measurement of phase $\phi$ is~\cite{Wiseman95b}
\begin{equation} \label{efF}
\hat F_{\mathrm{hom}} (x)dx =\frac{e^{-x^2/2} dx}{\sqrt{2\pi}} (|0 \rangle \langle 0| + x \hat{\sigma}_\phi + x^2|1 \rangle \langle 1 | ), 
\end{equation}
where 
\begin{equation}
\hat{\sigma}_\phi  = e^{i\phi}|0 \rangle \langle 1 | + e^{-i\phi}|1 \rangle \langle 0|.
\end{equation}
Binning over the sign of the result gives
\begin{equation} \label{Eff}
\hat E_\pm = \pm \int_0^{\pm \infty} \hat F_{\mathrm{hom}} (x)dx = \frac{1}{2} \left( \hat{1} \pm \sqrt{\frac{2}{\pi}} \hat{\sigma}_\phi \right)
\end{equation}
where $\hat 1$ is the identity operator in the qubit subspace. Thus the measurement is not projective in the qubit subspace even for 100\% efficient homodyne detection. Including inefficiency is equivalent to adding extra Gaussian noise to the measurement~\cite{WisMil10}, which is equivalent to replacing ${e^{-x^2/2} dx}{\sqrt{1/2\pi}}$ by ${e^{-\eta x^2/2} dx}{\sqrt{\eta/2\pi}} $. This changes Eq. (\ref{Eff}) to 
\begin{equation} \label{inEff}
\hat E_\pm = \frac{1}{2} \left( \hat{1} \pm \sqrt{\frac{2 \eta}{\pi}}\hat{\sigma}_\phi \right). 
\end{equation}

For Alice, it is irrelevant that the  the sign-binned homodyne measurement is not projective; all that matters for  the applicability of our theory is that she has a binary ($\pm1$-valued) measurement on a qubit. For Bob, it does matter that he is not performing a projective measurement. However, the positive-operator-valued measure (POVM) elements (\ref{inEff}) are simply  a mixture of the identity with the appropriate projectors, 
 \begin{equation}
 \hat\Pi_\pm = \frac{1}{2} \left( \hat{1} \pm \hat{\sigma}_\phi \right).
 \end{equation}
Thus any linear correlation between Alice's result $A$ and Bob's result $B$ is exactly the same as if Bob's POVM elements were these projectors, except that the correlations will be reduced by the 
 multiplicative factor 
 \begin{equation}
 \gamma = \sqrt{\frac{2\eta_{\rm Bob}}{\pi}}.
 \label{scaling_for_exp_imperfection}
 \end{equation}
 Thus, since Bob and his measurements are trusted in EPR-steering, we can apply our theory for projective measurements if we correct the measured correlations by dividing by $\gamma$.

\subsection{Analysis of the experimental results}

Now we calculate the EPR-steering inequality for Ref.~\cite{MorinPRL13} using the given experimental parameters [based on Eq. (\ref{eq:ineq_nec_suf})]. 
The only difference is that the left hand side of the EPR-steering inequality, which is linear in the correlation strength of Alice's and Bob's results, is reduced by the multiplicative factor of $\gamma$ [Eq. (\ref{scaling_for_exp_imperfection})]. Thus we can use (\ref{eq:ineq_nec_suf}) if we simply multiply the right hand side by the same factor $\gamma$, giving 
\begin{multline}
\sqrt{ \left\langle (A+A')B \right\rangle^2 + \left\langle (A+A')B' \right\rangle^2 } \\
+ \sqrt{ \left\langle (A-A')B \right\rangle^2 + \left\langle (A-A')B' \right\rangle^2 } \leq 2 \, \gamma. 
\label{cavalcanti_steering_correction}
\end{multline}
Here the observables denote sign-binned homodyne measurements for 
$ A \leftrightarrow  x , \, A' \leftrightarrow p, \,
B \leftrightarrow (x-p)/\sqrt{2}, \, B' \leftrightarrow (x+p)/\sqrt{2} $, 
where $x$ and $p$ denote the $x \> (\phi = 0)$ and $p \> (\phi = \pi /2)$ quadratures respectively. 

First we calculate the right hand side. The homodyne efficiency in Ref.~\cite{MorinPRL13} was $\eta_{\mathrm{Bob}} = 0.85$. Thus the right hand side evaluates to 
\begin{equation}
\mathrm{Right} = 2 \, \gamma = 2 \, \sqrt{\frac{2 \eta_{\mathrm{ Bob}}}{\pi}} = 1.47. 
\end{equation}

Next we calculate the left hand side. As Ref.~\cite{MorinPRL13} does not give individual correlation values, we assume that all correlations have the same magnitude: $ \langle AB \rangle = \langle AB' \rangle = \langle A'B \rangle = - \langle A'B' \rangle $, as expected from the experimental set-up, and as is usual for a CHSH-type correlation. With this assumption, the left hand side of Eq.~(\ref{cavalcanti_steering_correction}) reduces to
\begin{equation*}
\mathrm{Left} 
= 2 \times \sqrt{\left\langle AB \right\rangle^2 + \left\langle A' B' \right\rangle^2}
= 4 \langle AB \rangle.
\end{equation*}
whereas the $S$ value used to calculate the CHSH inequality in Ref.~\cite{MorinPRL13} reduces to 
\begin{equation*}
S \equiv \langle AB \rangle + \langle AB' \rangle + \langle A'B \rangle - \langle A'B' \rangle = 4  \langle AB \rangle = \mathrm{Left} . 
\end{equation*}
Since the largest $S$ value in Ref.~\cite{MorinPRL13} was $S_{\mathrm{max}} = 1.330$ for a maximally entangled state [$\theta = 22.5^\circ $ in Eq. (\ref{eq:model_arbitrary_qubit})], it follows that the EPR-steering inequality Eq. (\ref{cavalcanti_steering_correction}) is bounded by 
\begin{equation}
\mathrm{Left} \leq S_{\mathrm{max}} = 1.33  < 1.47 = \mathrm{Right}. 
\end{equation}
This proves that the experimentally measured correlations could not have violated an EPR-steering inequality.

\section{Conclusion}

We have derived an EPR-steering inequality and shown that its violation is necessary and sufficient to demonstrate EPR-steering in the simple scenario of 2 parties with 2 dichotomic measurements per party and mutually unbiased qubit observables for Bob. This inequality uses 4 correlations between the observables, as in the CHSH inequality, and we have shown that, as expected, the latter is violated only if our inequality is. We have then applied this inequality to a recent experiment involving a single photon split between two parties, and shown that this experiment could not have demonstrated EPR-steering, unlike a similar experiment by two of us and coworkers~\cite{Fuwa2015}, which did.

Questions for further research include: Is there a state which does not violate the inequality with mutually unbiased settings for Bob ($\mu=1/2$) but violates an EPR-steering inequality for other values of $\mu$? Or else, can we prove that violation of the inequality introduced here is necessary and sufficient to demonstrate steering under the scenario considered here? Can this procedure be adapted for more settings? And similar to the question above, are mutually unbiased settings optimal? Could it be sufficient (for two qubits) to consider a finite number of settings, i.e., is a 2-qubit state steerable if and only if it violates the necessary and sufficient inequality for, say, 3 settings? Otherwise, can we find a class of states that gives a counterexample?

\begin{acknowledgments}
This work was partly supported by the SCOPE program of the MIC of Japan, PDIS, GIA, G-COE, and APSA commissioned by the MEXT of Japan, FIRST initiated by the CSTP of Japan, ASCR-JSPS, and Australian Research Council grants CE110001027 and DE120100559. M.F. acknowledges financial support from ALPS. 
\end{acknowledgments}

\appendix

\section{Proof that all points inside $\mathcal{C}$ satisfy the LHV--LHS inequality} \label{appendix_A}

We first define the convex function $f(\mathbf{v}):=\sqrt{v_{1}^{2}+v_{2}^{2}}+\sqrt{v_{3}^{2}+v_{4}^{2}}$ for convenience. To see that $f$ is convex, simply note that it is the sum of two convex radial functions in $\mathbb{R}^{4}$. The statement that all points in $\mathcal{C}$ satisfy \eqref{eqn_ineq} may be formalised as: 
\begin{equation}
\mathbf{v}\in \mathcal{C}\implies f(\mathbf{v})\leq1.
\end{equation}
To prove this, first notice that any point in $\mathcal{C}$ is a convex combination of a single point from $\mathcal{C}_{1}$ and a single point from $\mathcal{C}_{2}$. Also notice that any point in $\mathcal{C}_{1}$ can be made up of a convex combination of two points on the bounding circle of $\mathcal{C}_{1}$ (similarly for $\mathcal{C}_{2}$). Therefore any point in $\mathbf{v}\in \mathcal{C}$ may be written as the convex combination: 
\begin{multline}
\mathbf{v}=p_{1}(\cos\xi_{1},\sin\xi_{1},0,0)+p_{2}(\cos\xi_{2},\sin\xi_{2},0,0)\\
+p_{3}(0,0,\cos\xi_{3},\sin\xi_{3})+p_{4}(0,0,\cos\xi_{4},\sin\xi_{4})
\end{multline}
where $\sum_{i}p_{i}=1$.

Expanding and simplifying the first two components of $\mathbf{v}$, we have 
\begin{align}
v_{1}^{2}+v_{2}^{2} & =p_{1}^{2}+p_{2}^{2}+2p_{1}p_{2}\cos(\xi_{1}-\xi_{2})\\
 & \leq p_{1}^{2}+p_{2}^{2}+2p_{1}p_{2}=(p_{1}+p_{2})^{2}
\end{align}
and similarly for $v_{3}^{2}+v_{4}^{2}$. Putting the components of $\mathbf{v}$ into the inequality \eqref{eqn_ineq} yields 
\begin{equation}
f(\mathbf{v})=\sqrt{v_{1}^{2}+v_{2}^{2}}+\sqrt{v_{3}^{2}+v_{4}^{2}}\leq p_{1}+p_{2}+p_{3}+p_{4}=1
\end{equation}
as desired.

Note that the inequality achieves the bound $1$ whenever $\xi_{1}=\xi_{2}$ and $\xi_{3}=\xi_{4}$, which is true whenever a point can be constructed
as a convex combination of a point on the boundary $\partial \mathcal{C}_{1}$ of $\mathcal{C}_{1}$ with a point on the boundary $\partial \mathcal{C}_{2}$ of $\mathcal{C}_{2}$. Here and in the following, the \emph{boundary of the set} $\mathcal{C}_{1}$ will mean the relative boundary of $\mathcal{C}_{1}$ with respect to the smallest affine subspace $\mathcal{P}_{1}$ in which $\mathcal{C}_{1}$ may be embedded.  Because $\mathcal{C}_{1}$
contains the origin this is in fact a linear subspace, and we may write $\mathcal{P}_{1}:=\mathrm{span}(\mathcal{C}_{1})$.  The relative boundaries $\partial \mathcal{C}_{1}$ and $\partial \mathcal{C}_{2}$ are circumferences, rather than boundaries with respect to the full space $\mathbb{R}^{4}$, which would be whole filled-in circles. In a similar manner, the interior of $\mathcal{C}_{1}$ denoted $\mathrm{int}(\mathcal{C}_{1})$ will mean the relative interior of $\mathcal{C}_{1}$ with respect to $\mathcal{C}_{1}$ rather than the full space.  See, for example, Ref.~\cite[\S 2.1]{Urruty2001}.

\section{Proof that only points inside $\mathcal{C}$ satisfy the LHV--LHS inequality}  \label{appendix_B}

Proving that 
\begin{equation}
f(\mathbf{v})\leq1\implies\mathbf{v}\in \mathcal{C}\label{eq:inequality sufficiency}
\end{equation}
is somewhat more complicated. The steps are:
\begin{enumerate}
\item Prove that $\mathbf{v}\in\partial \mathcal{C}\implies\mathbf{v}=p_{1}\mathbf{u}+p_{2}\mathbf{w}$ for some $\mathbf{u}\in\partial \mathcal{C}_{1}$ and $\mathbf{w}\in\partial \mathcal{C}_{2}$. The consequence of this is that the inequality reaches its bound of $1$ everywhere on the boundary of $\mathcal{C}$. In concise form, $f(\mathbf{v})=1\;\forall\mathbf{v}\in\partial \mathcal{C}$. 
\item Notice that $\mathcal{C}$ has an interior point, $\mathbf{0}\in\mathrm{int}(\mathcal{C})$ for which the convex function $f$ is strictly less than its value
on the boundary: $f(\mathbf{0})=0<1=f(\mathbf{v})\;\forall\mathbf{v}\in\partial \mathcal{C}$. In general, we prove that this implies that points \emph{not} in $\mathcal{C}$ do \emph{not} satisfy the inequality \eqref{eqn_ineq} and we have a proof of \eqref{eq:inequality sufficiency} by contradiction.
\end{enumerate}
Let us start with the second step above since it is simpler. We want to prove the following proposition:

\textbf{Proposition 1:} Suppose $f$ is a convex function and $\mathcal{C}$ is a closed set such that $f(\mathbf{c}_{i})=0$ for some $\mathbf{c}_{i}\in\mathrm{int}(\mathcal{C})$ while $f(\mathbf{c})=1$ for all $\mathbf{c}\in\partial \mathcal{C}$. Then $f(\mathbf{v})\leq1$ only if $\mathbf{v}\in \mathcal{C}$.

\emph{Proof.} Consider a point $\mathbf{w}\notin \mathcal{C}$; we prove that $f(\mathbf{w})>1$. Suppose for a contradiction that $f(\mathbf{w})\leq1$.
Since $\mathcal{C}$ is closed, there exists a point $\mathbf{c}_{b}\in\partial \mathcal{C}$ which is on the straight line between $\mathbf{c}_{i}$ and $\mathbf{w}$. We may write this point as the nontrivial convex combination $\mathbf{c}_{b}=p\mathbf{w}+(1-p)\mathbf{c}_{i}$ with $p<1$. By the convexity of $f$, we then have
\[
f(\mathbf{c}_{b})\leq(1-p)f(\mathbf{c}_{i})+pf(\mathbf{w})\leq p<1.
\]
However, this violates the assumption $f(\mathbf{c}_{b})=1$ and we therefore have $f(\mathbf{w})>1$. Thus, we have proven that $\mathbf{v}\notin \mathcal{C}\implies f(\mathbf{v})>1$, or equivalently, that $f(\mathbf{v})\leq1\implies\mathbf{v}\in \mathcal{C}$.

Given that we may take $\mathbf{c}_{i}=\mathbf{0}$, we are only left with proving that $f(\mathbf{v})=1$ for all points on the boundary of $\mathcal{C}$. For this we first prove two lemmas.

\textbf{Lemma 1:} Suppose $\mathcal{C}$ is a convex set and consider two points $\mathbf{c}\in\mathrm{int}(\mathcal{C})$ and $\mathbf{c}'\in \mathcal{C}$. Then any convex combination $\mathbf{c}''=(1-p)\mathbf{c}+p\mathbf{c}'$ with $p<1$ is in the interior of $\mathcal{C}$.

\emph{Proof.} This is Lemma 2.1.6 in \cite{Urruty2001}, but we prove it here for completeness. Since $\mathbf{c}\in\mathrm{int}(\mathcal{C})$, there exists a $\delta>0$ such that $\Vert\mathbf{e}-\mathbf{c}\Vert<\delta\implies\mathbf{e}\in \mathcal{C}$. Consider any point $\mathbf{e}''$ a small distance from $\mathbf{c}''$ such that $\Vert\mathbf{e}''-\mathbf{c}''\Vert<\delta(1-p)$. We have
\begin{align*}
\mathbf{e}'' & =(1-p)\mathbf{c}+p\mathbf{c}'+(\mathbf{e}''-\mathbf{c}'')\\
 & =(1-p)\left(\mathbf{c}+\frac{1}{1-p}(\mathbf{e}''-\mathbf{c}'')\right)+p\mathbf{c}'\\
 & =(1-p)\mathbf{e}+p\mathbf{c}'
\end{align*}
where we define $\mathbf{e}:=\mathbf{c+}\frac{1}{1-p}(\mathbf{e}''-\mathbf{c}'')$. Now notice that 
\[
\left\Vert \mathbf{e}-\mathbf{c}\right\Vert =\left\Vert \frac{1}{1-p}(\mathbf{e}''-\mathbf{c}'')\right\Vert <\delta
\]
because of the way that we chose $\Vert\mathbf{e}''-\mathbf{c}''\Vert$ --- this leaves us with $\mathbf{e}\in \mathcal{C}$. Therefore, we have $\mathbf{e}''$
as a convex combination of points $\mathbf{e},\mathbf{c}'\in \mathcal{C}$ which implies that $\mathbf{e}''$ is also in $\mathcal{C}$. In other words, $\Vert\mathbf{e}''-\mathbf{c}''\Vert<\delta(1-p)\implies\mathbf{e}''\in \mathcal{C}$ with the result that $\mathbf{c}''\in\mathrm{int}(\mathcal{C})$ as required. 

\textbf{Lemma 2}: Let $\mathcal{C}_{1}$ and $\mathcal{C}_{2}$ be closed convex sets which share a common interior point, $\mathbf{c}_{i}\in\mathrm{int}(\mathcal{C}_{1})\cap\mathrm{int}(\mathcal{C}_{2})$ and lie in orthogonal affine subspaces. Then $\mathrm{int}(\mathcal{C}_{1})\subset\mathrm{int}(\mathcal{C})$ where $\mathcal{C}=\mathrm{convex}(\mathcal{C}_{1}\cup \mathcal{C}_{2})$ is the convex combination of $\mathcal{C}_{1}$ and $\mathcal{C}_{2}$.

\emph{Proof.} Without loss of generality, we may assume that the shared interior point is the origin, $\mathbf{c}_{i}=\mathbf{0}$. We first need to prove that if $\mathbf{0}\in\mathrm{int}(\mathcal{C}_{1})\cap\mathrm{int}(\mathcal{C}_{2})$ then $\mathbf{0}\in\mathrm{int}(\mathcal{C})$.  What we need to show is that there exists $\delta>0$ such that for any point $\mathbf{u}'\in\mathrm{span}(\mathcal{C})$ a small distance from $\mathbf{0}$ such that with $\Vert\mathbf{u}'\Vert<\delta$, there exist points $\mathbf{u}_{1}\in \mathcal{C}_{1}$ and $\mathbf{u}_{2}\in \mathcal{C}_{2}$ such that $\mathbf{u}'$ can be written as a convex combination $\mathbf{u}'=p\mathbf{u}_{1}+(1-p)\mathbf{u}_{2}$ for some $\mu$. For this, denote the smallest subspaces containing $\mathcal{C}_{1}$ and $\mathcal{C}_{2}$ by $\mathcal{C}_{1}=\mathrm{span}(\mathcal{C}_{1})$ and $\mathcal{C}_{2}=\mathrm{span}(\mathcal{C}_{2})$,
respectively. Note that since $\mathcal{C}=\mathrm{convex}(\mathcal{C}_{1}\cup \mathcal{C}_{2})$, $\mathrm{span(}\mathcal{C})=\mathrm{span}(\mathrm{convex}(\mathcal{C}_{1}\cup \mathcal{C}_{2}))=\mathrm{span}(\mathcal{C}_{1}\cup \mathcal{C}_{2})=\mathrm{span}(\mathcal{C}_{1}\cup \mathcal{C}_{2})$. Therefore we can always write $\mathbf{u}'=\frac{1}{2}\mathbf{u}_{1}+\frac{1}{2}\mathbf{u}_{2}$ for some $\mathbf{u}_{1}\in \mathcal{C}_{1}$, $\mathbf{u}_{2}\in \mathcal{C}_{2}$. Since $\mathbf{0}$ is in the interior of both $\mathcal{C}_{1}$ and $\mathcal{C}_{2}$, there exist $\delta_1,\delta_2>0$ such that $\Vert\mathbf{u}_{1}\Vert<\delta_{1}\implies\mathbf{u}_{1}\in \mathcal{C}_{1}$ and $\Vert\mathbf{u}_{2}\Vert<\delta_{2}\implies\mathbf{u}_{2}\in \mathcal{C}_{2}$. At this point we exhibit the value of $\delta=\frac{1}{2}\min(\delta_{1},\delta_{2})$ which will do the trick. To show this, note that $\mathbf{u}_{1}$ and $\mathbf{u}_{2}$ are orthogonal which implies $\Vert\mathbf{u}_{1}\Vert\leq\Vert\mathbf{u}_{1}+\mathbf{u}_{2}\Vert$ and we have 
\begin{align*}
\Vert\mathbf{u}'\Vert & <\delta\\
\implies\Vert\mathbf{u}_{1}\Vert & \leq\Vert\mathbf{u}_{1}+\mathbf{u}_{2}\Vert=2\Vert\mathbf{u}'\Vert<2\delta=\min(\delta_{1},\delta_{2}) \le \delta_{1}\\
\implies\mathbf{u}_{1} & \in \mathcal{C}_{1}.
\end{align*}
Exactly the same holds for $\mathbf{u}_{2}$, and we have shown that $\mathbf{u}'\in \mathcal{C}$.

Now by Lemma 1, any convex combination $\mathbf{c}''=(1-p)\mathbf{0}+p\mathbf{c}'$ with $\mathbf{0}\in\mathrm{int}(\mathcal{C})$, $p<1$ and $\mathbf{c}'\in \mathcal{C}$
is in the interior of $\mathcal{C}$. However, all points in the interior of $\mathcal{C}_{1}$ (and $\mathcal{C}_{2}$) can be written in this form. To see this, imagine extending the ray between $\mathbf{0}$ and an arbitrary point $\mathbf{c}''\in \mathcal{C}_{1}$ to the boundary of $\mathcal{C}_{1}$ to get a point $\mathbf{c}_{b}\in\partial \mathcal{C}_{1}$, and simply take $\mathbf{c}'=\mathbf{c}_{b}$. Therefore $\mathrm{int}(\mathcal{C}_{1})\subset\mathrm{int}(\mathcal{C})$, as desired.

\textbf{Proposition 2:} Let $\mathcal{C}$ be the set defined by equation \eqref{eq:set_C_definition}. Then any point $\mathbf{c}\in\partial \mathcal{C}$ is a convex combination,
$\mathbf{c}=p\mathbf{c}_{1}+(1-p)\mathbf{c}_{2}$ of some points $\mathbf{c}_{1}\in\partial \mathcal{C}_{1}$ and $\mathbf{c}_{2}\in\partial \mathcal{C}_{2}$ where $\mathcal{C}_{1}$ and $\mathcal{C}_{2}$ are defined by equations \eqref{eq:sets_C1_C2_definitions}.

\emph{Proof}. Suppose for a contradiction that the proposition is false. Then there exists a $\mathbf{c}\in\partial \mathcal{C}$ such that any convex combination of points $\mathbf{c}_{1}\in \mathcal{C}_{1}$ and $\mathbf{c}_{2}\in \mathcal{C}_{2}$ with the property $p\mathbf{c}_{1}+(1-p)\mathbf{c}_{2}=\mathbf{c}$ has either (\emph{i}) $\mathbf{c}_{1}\in\mathrm{int}(\mathcal{C}_{1})$, $p>0$ or (\emph{ii}) $\mathbf{c}_{2}\in\mathrm{int}(\mathcal{C}_{2})$ and $p<1$. The situation is symmetric so it is enough to prove (\emph{ii}) to
be false. If (\emph{ii})\emph{ }is true then\emph{ }by Lemma 2 we have $\mathbf{c}_{2}\in\mathrm{int}(\mathcal{C})$. Therefore by Lemma 1 $\mathbf{c}\in\mathrm{int}(\mathcal{C})$
and we reach a contradiction.

This completes the proof of \eqref{eq:inequality sufficiency}.

\section{Maximum quantum violations of Inequality \eqref{eq:ineq_nec_suf}} \label{appendix_C}

We now consider quantum violations of inequality \eqref{eq:ineq_nec_suf}. Note first that \eqref{eq:ineq_nec_suf} can be rewritten as
\begin{multline}
\Bigl[\langle AB\rangle^2+\langle A'B\rangle^2+\langle AB'\rangle^2+\langle A'B'\rangle^2 \\
+ 2\left(\langle AB\rangle\langle A'B\rangle+\langle AB'\rangle\langle A'B'\rangle\right)\Bigr]^{\frac{1}{2}}\\
+ \Bigl[\langle AB\rangle^2+\langle A'B\rangle^2+\langle AB'\rangle^2+\langle A'B'\rangle^2 \\
- 2\left(\langle AB\rangle\langle A'B\rangle+\langle AB'\rangle\langle A'B'\rangle\right)\Bigr]^{\frac{1}{2}}.\label{eq:main_rewrite1}
\end{multline}

Suppose now Alice and Bob share a maximally entangled state of form
\begin{equation}\label{eq:max_ent}
|\Phi\rangle_{AB} =\frac{1}{\sqrt{2}} \left\{|1\rangle_A \otimes |1\rangle_B + |-1\rangle_A \otimes |-1\rangle_B\right\},
\end{equation}
where $\{|1\rangle_B, |-1\rangle_B\}$ are eigenstates of Bob's observable $B$ with eigenvalues $1$ and $-1$, respectively, and $\{|1\rangle_A, |-1\rangle_A\}$ are two orthogonal states for Alice's subsystem. Let Bob's observable $B'$ have eigenvalues $|1\rangle_B \pm |-1\rangle_B$ with respective eigenvalues $\pm1$ (thus satisfying the requirement of mutually unbiased measurements for Bob).

If Alice performs a projective measurement $A$ with outcome $a=1$ associated to the eigenstate $\mathrm{cos}(\alpha/2) |1\rangle_A + \mathrm{sin}(\alpha/2)|-1\rangle_A$, and outcome $a=-1$ associated to $\mathrm{sin}(\alpha/2) |1\rangle_A + \mathrm{cos}(\alpha/2)|-1\rangle_A$, then the correlation between $A$ and $B$ is $\langle AB\rangle=P(a=b)-P(a=-b)=\mathrm{cos}(\alpha)$, and that between $A$ and $B'$ is $\langle AB'\rangle=\mathrm{sin}(\alpha)$. Let the measurement $A'$ of Alice be similarly defined by $\alpha'$, and thus $\langle A'B\rangle=\mathrm{cos}(\alpha')$ and $\langle A'B'\rangle=\mathrm{sin}(\alpha')$.

With these choices of measurements, the left-hand side of inequality \ref{eq:main_rewrite1} becomes
\begin{multline}
\left[2 + 2\left(\mathrm{cos}(\alpha)\mathrm{cos}(\alpha')+\mathrm{sin}(\alpha)\mathrm{sin}(\alpha')\right)\right]^{\frac{1}{2}}\\
+\left[2 - 2\left(\mathrm{cos}(\alpha)\mathrm{cos}(\alpha')+\mathrm{sin}(\alpha)\mathrm{sin}(\alpha')\right)\right]^{\frac{1}{2}}\\
=\left[2 + 2\left(\mathrm{cos}(\alpha-\alpha')\right)\right]^{\frac{1}{2}} +\left[2 - 2\left(\mathrm{cos}(\alpha-\alpha')\right)\right]^{\frac{1}{2}}.\label{eq:main_rewrite2}
\end{multline}
This expression attains the maximum value of $2\sqrt{2}$ for $(\alpha-\alpha')=\pi/2$, which corresponds to mutually unbiased measurements $A$ and $A'$. Note that this does not depend on the absolute values of $\alpha$ or $\alpha'$, and thus does not depend on the relative angles between Alice's and Bob's measurements, unlike the case of the CHSH inequality which does require specific angles between Alice and Bob in order to maximise the violation for a given maximally entangled state.

It seems intuitive that allowing for POVMs or non-maximally entangled states should not increase the violation of the inequality, but we leave this as an open question.

\end{document}